\documentclass[conference]{IEEEtran}

\usepackage{algorithm}
\usepackage{algpseudocode}

\usepackage[acronym]{glossaries}
\usepackage{graphicx}

\usepackage{listings}
\usepackage{xcolor}

\usepackage{xurl}
\usepackage{xspace}
\usepackage{lscape}
\usepackage{float}
\usepackage{tabularx}
\usepackage{booktabs}
\usepackage{amsmath}

\usepackage{relsize} 
\newcommand{\class}[1]{\textsc{#1}}
\newcommand{\func}[1]{\textit{#1}}         
\newcommand{\code}[1]{\texttt{\smaller #1}} 

\definecolor{codegreen}{rgb}{0,0.6,0}
\definecolor{codegray}{rgb}{0.5,0.5,0.5}
\definecolor{codepurple}{rgb}{0.58,0,0.82}
\definecolor{backcolour}{rgb}{0.95,0.95,0.92}

\lstdefinestyle{mystyle}{
    backgroundcolor=\color{backcolour},   
    keywordstyle=\color{blue},
    commentstyle=\color{gray},
    stringstyle=\color{red},
    numberstyle=\tiny\color{codegray},
    basicstyle=\ttfamily\footnotesize,
    breakatwhitespace=false,         
    breaklines=true,                 
    captionpos=b,                    
    keepspaces=true,                 
    numbers=left,                    
    numbersep=5pt,                  
    showspaces=false,                
    showstringspaces=false,
    showtabs=false,                  
    tabsize=2
}

\begin{document}

\title{Simulating Dynamic Cloud Marketspaces:\\ Modeling Spot
Instance Behavior and Scheduling with CloudSim Plus}


\author{\IEEEauthorblockN{Christoph Goldgruber\\}
\IEEEauthorblockA{\textit{Faculty of Computer Science} \\
\textit{University of Vienna}\\
Vienna, Austria\\
christoph.goldgruber@univie.ac.at}
\and
\IEEEauthorblockN{Benedikt Pittl\\}
\IEEEauthorblockA{\textit{Faculty of Computer Science} \\
\textit{University of Vienna}\\
Vienna, Austria \\
benedikt.pittl@univie.ac.at}
\and
\IEEEauthorblockN{Erich Schikuta\\}
\IEEEauthorblockA{\textit{Faculty of Computer Science} \\
\textit{University of Vienna}\\
Vienna, Austria \\
erich.schikuta@univie.ac.at}
}

\maketitle

\begin{abstract}
        The increasing reliance on dynamic pricing models, such as spot instances, in public cloud environments presents new challenges for workload scheduling and reliability. While these models offer cost advantages, they introduce volatility and uncertainty that are not fully addressed by current allocation algorithms or simulation tools.
        This work contributes to the modeling and evaluation of such environments by extending the CloudSim Plus simulation framework to support realistic spot instance lifecycle management, including interruption, termination, hibernation, and reallocation. The enhanced simulator is validated using synthetic scenarios and large-scale simulations based on the Google Cluster Trace dataset.
        Building on this foundation, the HLEM-VMP allocation algorithm—originally proposed in earlier research—was adapted to operate under dynamic spot market conditions. Its performance was evaluated against baseline allocation strategies to assess its efficiency and resilience in volatile workload environments. The comparison demonstrated a reduction in the number of spot instance interruptions as well as a decrease in the maximum interruption duration.
        Overall, this work provides both a simulation framework for simulating dynamic cloud behavior and analytical insights into virtual machine allocation performance and market risk, contributing to more robust and cost-effective resource management in cloud computing.
\end{abstract}


\IEEEpeerreviewmaketitle

\section{Introduction}
\label{intro}

The concept of delivering computing resources as a public utility was proposed by John McCarthy in 1961 and gradually realized in the late 1990s \cite{surbiryala2019cloud}. Cloud computing gained mainstream adoption around 2007, driven in part by initiatives from Google and IBM to provide cloud platforms to universities worldwide \cite{namasudra2017cloud}. Today, it underpins major technologies such as big data, IoT, and mobile Internet, generating over \$330 billion in annual revenue \cite{cloudmarket}.

The National Institute of Standards and Technology (NIST) defines cloud computing as on-demand network access to a shared pool of configurable resources that can be provisioned and released with minimal management effort \cite{mell2011nist}. Its core characteristics include on-demand self-service, broad network access, resource pooling, rapid elasticity, and measured service. Common service models are Software as a Service (SaaS), Platform as a Service (PaaS), and Infrastructure as a Service (IaaS), offered through deployment models such as public, private, community, and hybrid clouds \cite{mell2011nist, antonopoulos2017cloud}.

Cloud computing offers cost-efficiency, scalability, and flexibility compared to on-premise infrastructure but raises concerns around data security, compliance, and multi-tenancy risks \cite{singh2017cloud}. These challenges are particularly relevant in the IaaS model, where users manage virtual machines (VMs) but depend on providers for the underlying infrastructure.

The broad adoption of cloud platforms has intensified research on resource allocation, scheduling, energy efficiency, and VM migration policies \cite{rahman2019nutshell}. Testing such solutions directly on public clouds is costly and lacks full control over internal behavior, motivating the use of simulation for reproducible, large-scale experimentation \cite{calheiros2011cloudsim}.

This work extends the CloudSim Plus simulation framework to accurately model \emph{spot instances}—discounted, preemptible VMs reclaimed by providers when capacity is needed. Spot instances offer significant cost savings but introduce volatility that existing frameworks do not natively support. The proposed extension models full spot instance lifecycles, including termination, hibernation, and dynamic reprovisioning, enabling evaluation of interruption-aware allocation strategies.

The remainder of this paper is organized as follows: Section~\ref{stateofart} reviews enabling technologies and simulation frameworks. Section~\ref{question} presents research questions and objectives. Section~\ref{requirements} defines system requirements. Section~\ref{implementation} describes the simulation extension. Section~\ref{algorithm} introduces and adapts an existing VM allocation algorithm. Section~\ref{evaluation} evaluates the proposed solution using synthetic and real workload traces.

\section{State of the Art}
\label{stateofart}

This section reviews key enabling technologies for Infrastructure as a Service, common purchasing models, and existing cloud simulation tools relevant to modeling spot instance behavior.

\subsection{Enabling Technologies}

\emph{Virtualization} allows multiple isolated VMs to run on shared physical hardware through a hypervisor \cite{asvija2019security}. Type~1 hypervisors (e.g., Xen, VMware ESXi) run directly on hardware, offering lower overhead; Type~2 hypervisors (e.g., VirtualBox) run atop a host OS with added flexibility but reduced performance \cite{portnoy2012virtualization}. Full virtualization runs unmodified guest OSes, while paravirtualization requires OS modifications for improved efficiency \cite{rashid2019virtualization}.

\emph{Containerization} offers lighter-weight isolation at the OS level \cite{bhardwaj2021virtualization}, enabling faster startup and higher density than VMs but with less OS flexibility. Orchestration systems such as Kubernetes support scalability and fault tolerance.

\subsection{Common Purchase Models} \label{common}

Virtual machine instances are generally offered through several widely adopted purchase models. While cloud providers may use different terminology, this section follows the conventions used by Amazon Web Services (AWS), the market leader \cite{cloudmarket}. The three principal models are \emph{On-Demand Instances}, \emph{Reserved Instances}, and \emph{Spot Instances}. In addition, some providers also offer \emph{Dedicated Hosts}, which grant exclusive use of physical servers.

\paragraph{On-Demand Instances.}

On-demand instances represent the default and most flexible consumption model for virtual machines \cite{awsdocs}. They operate on a \emph{pay-as-you-go} basis, without requiring long-term commitments or upfront payments. Users are billed only for the duration the instance is running \cite{deldari2020survey}, and most providers now offer \emph{per-second billing} with a minimum granularity of 60 seconds \cite{googledocs, awsdocs}.  

The lifecycle of on-demand instances is fully user-controlled: instances can be started, stopped, or terminated at any time. This flexibility makes them attractive for unpredictable or short-lived workloads, such as testing environments, prototypes, or applications with sudden traffic spikes \cite{azuredocs}. However, the convenience comes at a premium—on-demand instances are generally the most expensive option in terms of unit price.

\paragraph{Reserved Instances.}

Reserved instances require users to commit to a fixed contract period, typically one or three years \cite{awsdocs, azuredocs, googledocs}. In exchange, they provide significant cost savings—up to 72\% compared to on-demand pricing—making them suitable for predictable, steady-state workloads such as web applications or backend services.  

Payment models vary: users may choose full upfront, partial upfront, or monthly installments. The key challenge is forecasting demand: underestimating may lead to on-demand overflow costs, while overestimating results in underutilized reservations and wasted expenditure \cite{ambati2020no}.  

To reduce risk, some providers allow limited modifications (e.g., instance size, family, or availability zone). Azure even permits early termination with partial refunds, albeit with fees \cite{azuredocs}. AWS additionally supports a \emph{Reserved Instance Marketplace}, where users can resell unused reservations. While resale provides flexibility, it is not frictionless: AWS charges a 12\% transaction fee, and only the upfront portion of reservations is transferrable \cite{yang2019online}. Research has proposed algorithmic trading strategies to optimize resale decisions \cite{yang2019online}.  

\paragraph{Spot / Preemptible Instances.} \label{spot}

Spot or preemptible instances utilize excess capacity in cloud data centers and are offered at discounts of up to 90\% compared to on-demand pricing \cite{awsdocs, george2019analyzing}. The trade-off is reliability: providers may revoke instances at any time, typically with a short warning period (e.g., 30 seconds on GCP, two minutes on AWS).  

Originally, spot pricing followed a pure market-based auction, producing highly volatile prices that encouraged research into bidding strategies \cite{baughman2019deconstructing}. In 2017, AWS switched to a more stable pricing scheme, likely based on smoothed demand–supply trends, though the exact mechanism remains undisclosed. Volatility decreased and long-term averages dropped, but studies found that short-lived workloads became relatively more expensive, while long-term spot usage benefited from the stability \cite{baughman2019deconstructing, george2019analyzing}.  

Today, spot instances are most effective for \emph{fault-tolerant, parallelizable, and time-flexible workloads}, such as data analytics, rendering, or large-scale simulations \cite{azuredocs, googledocs}. They are also commonly used in research contexts to reduce cost, provided interruptions can be tolerated. AWS provides historical price trends and instance interruption frequencies to support planning \cite{awsconsole}, but the inherent uncertainty makes workload adaptation strategies—such as checkpointing or hibernation—essential. Their operational lifecycle is discussed further in Section~\ref{lifecycle}.

\paragraph{Dedicated Hosts.}

While not a purchase model in themselves, dedicated hosts grant exclusive access to a physical server. They can be billed either on-demand or reserved, depending on contract length \cite{awsdocs, azuredocs}. This option is primarily aimed at enterprises with strict compliance, licensing, or performance isolation requirements. For instance, dedicated hosts allow customers to bring existing per-socket or per-core software licenses, or to ensure regulatory data segregation. The trade-off is cost, as dedicated hosts are substantially more expensive than multi-tenant instances.

\subsection{Cloud Simulation}
\label{simulation}

Evaluating cloud resource management strategies directly on production infrastructure is costly, time-consuming, and lacks reproducibility. Cloud simulation addresses these challenges by providing a controlled, repeatable, and scalable environment to test algorithms under diverse workloads and configurations \cite{calheiros2011cloudsim, law2013simmodeling}. It allows researchers to model datacenter architectures, virtualized resources, and scheduling policies without requiring access to expensive infrastructure.

\begin{figure}[!t]
    \centering
    \includegraphics[width=\linewidth]{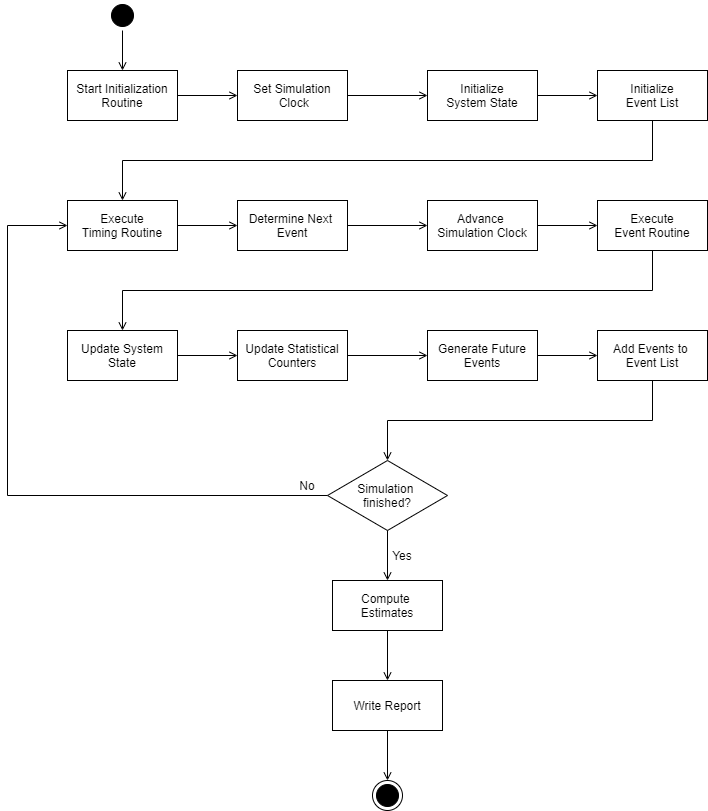}
    \caption{Execution flow of a discrete event simulation~\cite{law2013simmodeling}.}
    \label{fig:desflow}
\end{figure}

Most cloud simulation frameworks employ \emph{discrete event simulation} (DES), where the system state evolves at discrete event times rather than continuously. DES enables fine-grained modeling of resource allocation, job arrivals, and VM lifecycle events. Figure~\ref{fig:desflow} illustrates the typical execution flow of a DES model, consisting of entities (e.g., datacenters, brokers, VMs), resources, events, and a central event queue. Simulation time can be advanced either through a next-event mechanism or fixed time increments, with the former being more common in cloud simulators due to efficiency.

Several frameworks have been developed for cloud research. The most widely used is \emph{CloudSim}, introduced by Calheiros et al.~\cite{calheiros2011cloudsim}. It provides models for datacenters, physical hosts, VMs, network bandwidth, and workloads, and supports extensibility through custom policies. Building on this foundation, \emph{CloudSim Plus} is a modern fork that improves modularity, maintainability, and performance. It is actively maintained, making it suitable for research requiring extensions or large-scale experimentation. Numerous derivatives have been built on CloudSim, such as WorkflowSim (workflow scheduling), NetworkCloudSim (network modeling), and iFogSim (fog/edge computing).

Alternative simulators include \emph{SimGrid}, originally developed for distributed systems; \emph{iCanCloud}, focusing on scalability and energy modeling; and \emph{GreenCloud}, tailored to power consumption analysis. Each has specific strengths, but CloudSim and CloudSim Plus dominate academic usage due to their generality and community adoption.

Despite these strengths, a notable limitation of existing frameworks is the lack of native support for \emph{spot or preemptible instances}. These discounted VMs play a central role in modern cloud economics, yet their interruption-prone nature is not accurately captured. This limits the ability to evaluate scheduling policies that must account for uncertainty in instance lifetimes. Extending CloudSim Plus to model spot instance behavior directly addresses this gap and enables more realistic experimentation in IaaS environments.

\subsection{VM Allocation Algorithms}
\label{algorithms}

Virtual machine allocation is a core problem in cloud resource management, as it determines how virtualized workloads are mapped onto physical hosts. The objective is to maximize utilization of resources such as CPU, memory, and bandwidth, while minimizing energy consumption, costs, and violations of service-level agreements \cite{masdari2016overview, attaoui2024multi}. The design of allocation policies thus has a direct impact on both provider efficiency and user experience.

Allocation strategies are commonly divided into \emph{static} and \emph{dynamic} approaches. In static allocation, VMs remain on the host initially selected until termination. While this reduces overhead and complexity, it lacks adaptability to workload fluctuations and can lead to inefficient resource usage. In contrast, dynamic allocation allows VMs to be migrated during runtime, enabling load balancing, power-aware consolidation, and SLA compliance at the expense of additional migration cost.

A wide range of algorithms have been proposed. \emph{Heuristic-based methods}, such as First Fit, Next Fit, Best Fit, and Worst Fit, provide lightweight solutions with low computational overhead \cite{gupta2014survey}. These algorithms are easy to implement and scale well to large infrastructures, but they often fail to achieve globally optimal results. In contrast, \emph{metaheuristic approaches}, including genetic algorithms, ant colony optimization, and swarm-based techniques, explore a larger solution space and can deliver higher-quality placements \cite{mehta2024improved, murali2023meta, ciptaningtyas2023genvpm}. Their drawback lies in high execution time and complexity, which can limit applicability in environments requiring rapid decisions.

A further category is formed by \emph{constraint-based and weighted scoring algorithms}, which attempt to balance solution quality with computational efficiency. These approaches evaluate candidate hosts across multiple resource dimensions, such as CPU, memory, and bandwidth capacities, typically by applying scoring functions or cost models. An example is the \emph{HLEM-VMP algorithm} (Heuristic-based Load balancing and Energy-aware VM Placement), which combines host filtering with weighted evaluation of feasible hosts to minimize overload and reduce SLA violations \cite{jing2024hlemvmp}. Such algorithms achieve higher placement quality than simple heuristics while avoiding the significant computational overhead of metaheuristics, making them particularly suitable for large-scale simulation studies.

\section{Research Questions}
\label{question}

This section presents the research objectives and questions addressed in this study.  
The focus lies on simulating spot instance behavior and evaluating the impact of optimized allocation algorithms under dynamic cloud conditions.  
The central research questions are defined as follows.

\begin{table}[H]
\caption{Research questions}
\label{tab:research_question}
\centering
\begin{tabularx}{\columnwidth}{l X}
\hline
\textbf{ID} & \textbf{Research question} \\
\hline
\textbf{RQ1} &
How can spot instances be accurately modeled and simulated using cloud simulation tools? \\
\textbf{RQ2} &
How can virtual machine allocation algorithms be optimized to reduce the likelihood of spot instance interruptions through informed host selection in cloud simulation environments? \\
\hline
\end{tabularx}
\end{table}

To address \emph{RQ1}, this work investigates how to effectively simulate the key characteristics of spot instances. This includes:
\begin{itemize}
    \item modeling spot instance behavior within an established cloud simulation framework;
    \item accurately simulating interruption behaviors, including termination and hibernation;
    \item extending virtual machine lifecycle states to support dynamic provisioning and resubmission;
    \item utilizing real-world trace data as a proof of concept for validating spot behavior modeling and assessing allocation efficiency;
    \item identifying factors contributing to interruptions, and examining whether specific instance characteristics can be used to model interruption probabilities.
\end{itemize}

For \emph{RQ2}, the research focuses on enhancing virtual machine allocation strategies. Specifically, it explores how cloud simulation tools can evaluate and improve algorithms under realistic scenarios involving spot instances. Key tasks include:
\begin{itemize}
    \item adapting suitable allocation algorithms from existing literature;
    \item introducing deterministic modifications to improve the algorithm's ability to handle spot instance interruptions, with the goal of reducing both their frequency and duration;
    \item conducting performance comparisons between the improved algorithm and existing allocation methods in scenarios that involve both spot and on-demand instances.
\end{itemize}

Collectively, these contributions seek to increase the fidelity of cloud simulation environments by integrating realistic resource models commonly utilized by cloud providers. The resulting simulation framework allows researchers and practitioners to evaluate virtual machine allocation strategies under realistic constraints, providing insights into cost–performance trade-offs within volatile cloud markets.

\section{Requirement Analysis}
\label{requirements}

The goal of this section is to identify the features necessary for the implementation. The primary purpose is to improve understanding of how spot instances behave in a cloud environment and to derive the corresponding requirements for modeling and implementing this behavior. The section begins with an introduction to the general cloud architecture, followed by a discussion of the lifecycle and behavior of spot instances. Functional and non-functional requirements are presented in Sections~\ref{functional} and~\ref{nonfunctional}, respectively.

\subsection{Background}

To accurately simulate the behavior of spot instances within a cloud environment, the first step is to identify the essential entities in the cloud architecture. The National Institute of Standards and Technology defines five major actors in its cloud computing reference architecture: the cloud provider, cloud consumer, cloud broker, cloud auditor, and cloud carrier~\cite{liu2011nist}.

\noindent\textbf{Cloud provider:}  
Organizations that offer cloud services and resources to other parties. They manage the physical infrastructure (servers, storage, networking) and software stack (hypervisors, host operating systems) needed for virtualization and provisioning~\cite{liu2011nist}. Service models and key market providers are discussed in Sections~\ref{intro} and~\ref{stateofart}.

\noindent\textbf{Cloud consumer:}  
Users or organizations that acquire cloud resources and services from providers. In IaaS, consumers can deploy virtual machines, use storage, and access compute resources to run applications. Pricing is generally based on the quantity and duration of consumed resources. Terms are defined in service level agreements~\cite{liu2011nist}.

\noindent\textbf{Cloud broker:}  
Acts as an intermediary between consumers and providers, offering value-added services such as identity management or aggregating multiple providers’ services into a unified offering~\cite{liu2011nist}.

\noindent\textbf{Cloud auditor:}  
Independently assesses cloud services for security, privacy, and performance, ensuring compliance with regulations~\cite{liu2011nist}.

\noindent\textbf{Cloud carrier:}  
Facilitates network connectivity between providers and consumers, typically offering telecommunication and data transport services~\cite{liu2011nist}.

For this work, the relevant actors for simulating a cloud environment are:  
1) the \emph{cloud provider},  
2) the \emph{cloud consumer}, and  
3) the \emph{cloud broker}.

\subsubsection{Cloud Datacenters}

The foundation of the cloud infrastructure lies in datacenters. Major providers operate multiple datacenters across global regions. Infrastructure includes computing resources, storage systems, networking, and supporting systems for power and cooling. Large-scale datacenters can span multiple buildings with thousands of servers~\cite{datacentergoogle}.

Virtualization enables abstraction of computing, storage, and networking~\cite{moreno2012iaas}. Moreno et al.~\cite{moreno2012iaas} describe a reference architecture where the \emph{Cloud OS} manages virtual resources in multi-tenant environments. It handles VM lifecycle operations (deployment, migration, suspension) via a virtual machine manager, monitors resources through an information manager, and manages applications across VMs through a service manager. A scheduler allocates VMs to hosts based on capacity and requirements.

Other components include network management, accounting and auditing, authentication and authorization, storage/image management, and administrative tools~\cite{moreno2012iaas}.

\subsubsection{Virtual Machine Placement}

VM placement assigns virtual machines to physical hosts in a way that balances service quality, energy efficiency, and capacity utilization~\cite{mann2015allocation, silva2018approaches}. VMs are characterized by CPU cores, MIPS per core, memory, and disk size~\cite{mann2015allocation}.

Placement must respect host resource limits, adapt to changing workloads, and meet SLA requirements. Failure incurs operational costs~\cite{mann2015allocation}. While numerous placement strategies exist~\cite{silva2018approaches, masdari2016overview}, this work focuses on spot instance behavior, discussed next.

\subsubsection{Spot Instance Lifecycle}
\label{lifecycle}

Pham et al.~\cite{pham2018performance} define a lifecycle model for AWS EC2 Spot Instances. A spot request includes a machine type, maximum bid, and persistence option~\cite{awsconsole}. Requests can be:
1) unfulfilled,
2) fulfilled with interruption, or
3) fulfilled without interruption.

Unfulfilled requests may be cancelled by the user, expire, or be rejected for low bid or capacity limits. Fulfilled instances may later be interrupted (due to insufficient capacity or bid price below market rate) or terminated by the user. AWS provides a \emph{warning time}—about two minutes—before interruption~\cite{pham2018performance}.

\noindent\textbf{Interruption behavior:}  
Interruptions are mainly due to capacity reallocation to on-demand/reserved instances or low bid prices~\cite{pham2018performance}. Experiments found that most requests (80–90\%) are fulfilled within 4~s; unfulfilled ones often wait $>$60~s. No correlation was found between waiting time, workload intensity, or bid value and interruption frequency. Price stability changes in AWS have reduced price-based interruptions. AWS publishes interruption frequency data by instance type~\cite{awsspotadv}.

\begin{figure}[!t]
    \centering
    \includegraphics[width=\linewidth]{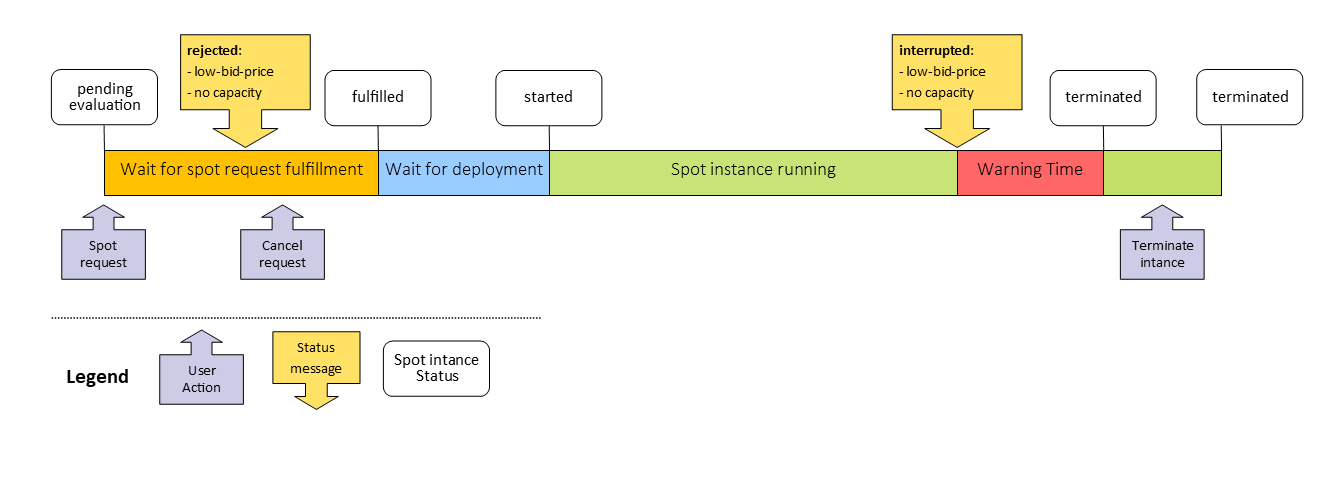}
    \caption{Spot instance lifecycle based on EC2, showing events, status messages, and user actions~\cite{pham2018performance}.}
    \label{fig:lifecycle}
\end{figure}

Based on these models, the following requirements are defined.

\subsection{Functional Requirements}
\label{functional}

Essential capabilities for realistic spot instance modeling in simulation:

\begin{itemize}
    \item \textbf{VM allocation behavior:} Spot instances must be interruptible at any time due to on-demand capacity demands.
    \item \textbf{VM instance types:} The framework must support distinct VM types (e.g., on-demand, spot) with different runtime behaviors.
    \item \textbf{VM lifecycle extensions:} Unallocated VMs should be retained and eligible for resubmission when resources become available.
    \item \textbf{Interruption handling:} Interruptions result in either hibernation or termination; a configurable minimum runtime must be enforced before interruptions.
    \item \textbf{Resubmission logic:} Hibernated instances must be resubmitted periodically, and unallocated requests retained for a configurable timeout.
    \item \textbf{Simulation output and monitoring:} The framework must log execution history, interruption counts, and average interruption times for post-simulation evaluation.
\end{itemize}

\noindent\textbf{Rationale:}  
Most frameworks model only on-demand instances. This implementation extends them to support spot-specific features such as configurable interruption and hibernation, minimum runtime guarantees, resubmission logic, and detailed execution metrics.

\subsection{Non-Functional Requirements}
\label{nonfunctional}

Requirements for the simulation framework:

\begin{itemize}
    \item \textbf{Modularity:} Architecture must allow isolated development of new features.
    \item \textbf{Extensibility:} New functionality should integrate without invasive code changes, including advanced allocation algorithms.
    \item \textbf{Maintainability:} Code must be logically organized with clear abstractions.
    \item \textbf{Documentation and transparency:} The framework must be open-source and well-documented.
    \item \textbf{Scientific relevance:} The framework should be widely recognized in research, supporting reproducibility and comparability.
    \item \textbf{Configurability:} The spot instance model must allow user-defined parameters (e.g., minimum runtime, hibernation timeout).
    \item \textbf{Observability:} Mechanisms must exist to track and log spot instance state changes, interruptions, and resubmissions.
\end{itemize}

\noindent\textbf{Rationale:}  
To model dynamic, interruption-prone workloads, the framework must be open-source, modular, and extensible. It should support advanced allocation logic, maintainable code, and provide observability for evaluation.

\section{Modeling Spot Instances in CloudSim Plus}  
\label{implementation}

This section describes the modeling and implementation of a CloudSim Plus extension for modeling spot instances and their dynamic behavior. The extension introduces new classes and mechanisms to capture spot lifecycle states, interruptions, and resubmission policies. The implementation builds on CloudSim Plus version~6.2.10.\footnote{Source code: \url{https://github.com/CGoldi/cloudsimplus6-spot-instance.git}.}

CloudSim Plus is an actively maintained fork of CloudSim 3, addressing major shortcomings of its predecessor, such as poor code quality, lack of design patterns, and extensive code duplication~\cite{silva2017cloudsim}. It reduced duplicated code lines by 30.44\%, whereas CloudSim 4 exhibited an increase of 300.96\%~\cite{silva2017cloudsim}. These improvements enhanced maintainability, testability, and extensibility. Additional advantages include integration testing, comprehensive documentation, and performance improvements across VM allocation policies~\cite{cloudsimplusdocs, cloudsimplusweb}.  

\subsection{Core Simulation Components of CloudSim Plus}

The CloudSim Plus engine is built around a set of core classes that manage event scheduling and enable communication between simulation entities. Together, these components provide the execution backbone of the simulator.

\paragraph{CloudSim.} 
The \class{CloudSim} class is the central entry point of the framework. It initializes the simulation, advances the internal clock, and processes scheduled events until either all events are executed or a predefined termination condition is reached. Events are stored in a future event queue sorted by timestamp and transferred to a deferred queue when they are due for processing \cite{calheiros2011cloudsim}.

\paragraph{SimEntity.} 
The \class{SimEntity} interface defines the contract for all simulation participants. Entities can schedule, send, and receive events, thereby driving the simulation forward. All entities extend the abstract class \class{CloudSimEntity}, which provides common logic for event handling. Typical examples are the \class{Datacenter}, \class{DatacenterBroker}, and \class{CloudletScheduler}.

\paragraph{SimEvent.} 
The \class{SimEvent} class represents the structure of a simulation event. Each event contains a type identifier, timestamp, source and destination entities, and an optional payload for passing additional information. Events therefore act as the communication mechanism between simulation components.

\paragraph{CloudSimTags.} 
The \code{CloudSimTags} class defines static constants used to identify events. These tags guide how an event is interpreted by the receiving entity. Examples include \code{END\_OF\_SIMULATION}, \code{CLOUDLET\_SUBMIT}, and \code{VM\_DESTROY}. Tags ensure that events are processed consistently across all entities.

In combination, these components provide a modular event-driven foundation that supports the extensibility of CloudSim Plus and enables researchers to model a wide variety of cloud behaviors.

\subsection{Key Infrastructure Classes in CloudSim Plus}

Beyond the simulation kernel, CloudSim Plus provides a set of infrastructure classes that define the system model. These classes represent physical resources, virtual resources, and user-facing entities.

\paragraph{DatacenterSimple.} 
This class models a virtualized datacenter as a collection of \class{Host} objects. Each datacenter is associated with a VM allocation policy that determines how incoming VMs are mapped to hosts.

\paragraph{VmAllocationPolicyAbstract.} 
An abstract base for implementing VM placement and migration strategies. Developers can override its default behavior (e.g., \func{defaultFindHostForVm()}) to implement custom allocation heuristics. Several policies are included by default, such as \emph{FirstFit}, \emph{BestFit}, and \emph{RoundRobin}.

\paragraph{HostSimple.} 
Represents a physical server with configurable CPU cores (PEs), memory, bandwidth, and storage. A host can run multiple VMs and apply either space-shared or time-shared policies for CPU scheduling.

\paragraph{DatacenterBrokerAbstract.} 
Acts as a user-side agent managing the lifecycle of VMs and cloudlets. The broker can be extended to implement different strategies for submitting and scheduling workloads, making it one of the main customization points for researchers.

\paragraph{VmSimple.} 
Defines the virtual machine abstraction, with configurable resources such as RAM, storage, bandwidth, and PEs. VMs execute cloudlets and use a \class{CloudletScheduler} to manage task execution.

\paragraph{Cloudlet.} 
Represents an application task or job. A cloudlet is characterized by its required PEs and number of instructions (in MIPS). Resource usage models allow workloads to consume CPU, memory, and bandwidth in different ways.

\paragraph{EventListener.} 
Listeners enable runtime monitoring by triggering actions when specific events occur, such as VM allocation, cloudlet completion, or simulation time progression. They provide a flexible mechanism for extending the simulator without modifying its core logic.

These infrastructure classes, together with the simulation kernel, provide the structural foundation upon which extensions—such as spot instance modeling—can be implemented and evaluated \cite{cloudsimplusdocs, silva2017cloudsim}.

            \subsection{Modeling Objectives}

            This section introduces the modeling objectives of the spot instance extension and illustrates how the design abstracts spot instance behavior within the CloudSim Plus simulation framework. The aim is to conceptually map the new functionality onto the architecture of the existing simulation environment and to explain how the new components interact with core elements such as the virtual machine lifecycle, allocation policies, and simulation control flow.
            
\begin{figure}[!t]
    \centering
    \includegraphics[width=\columnwidth]{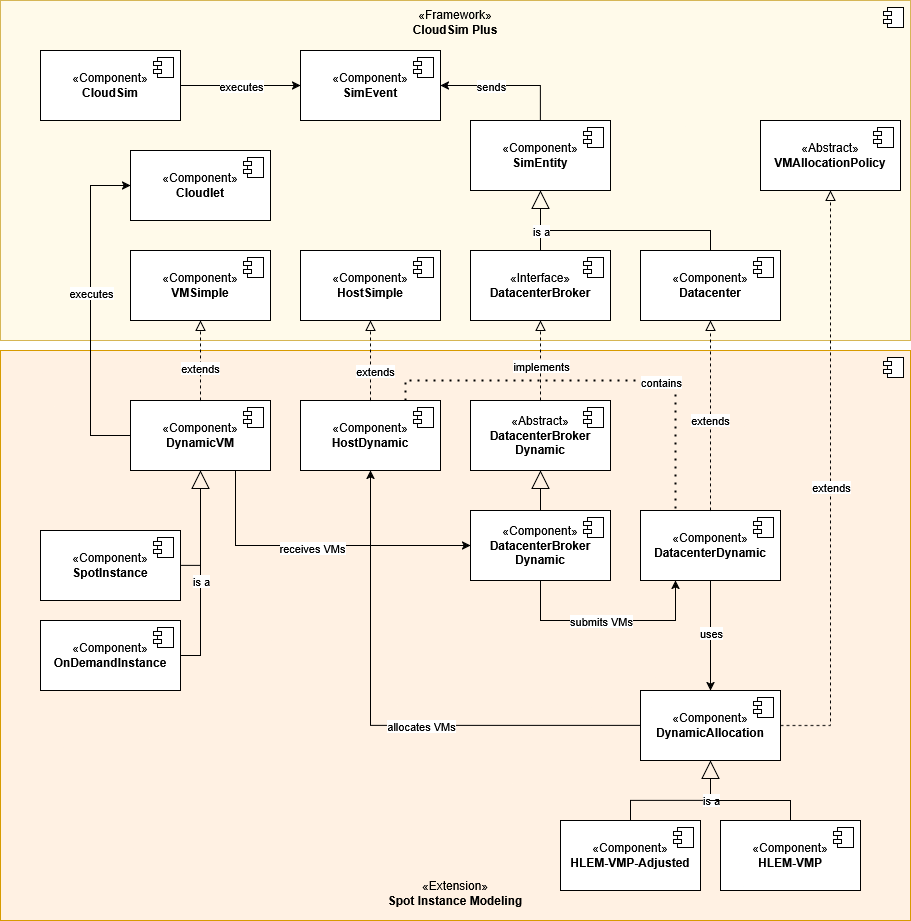}
    \caption{Component diagram of the spot instance modeling extension. Yellow: original modules; orange: new components.}
    \label{fig:spot_component_diagram}
\end{figure}

            Figure~\ref{fig:spot_component_diagram} provides a high-level component view of the integration. The yellow box highlights the original CloudSim Plus modules, while the orange box labeled “Spot Instance Modeling” encapsulates the new components introduced by the extension. These components implement core functionality required to simulate dynamic spot instance behavior, including lifecycle transitions, interruption logic, and resubmission mechanisms. The diagram also reflects how spot-specific components interface with host scheduling and VM allocation policies during simulation execution.
            
\begin{figure}[!t]
    \centering
    \includegraphics[width=0.70\columnwidth]{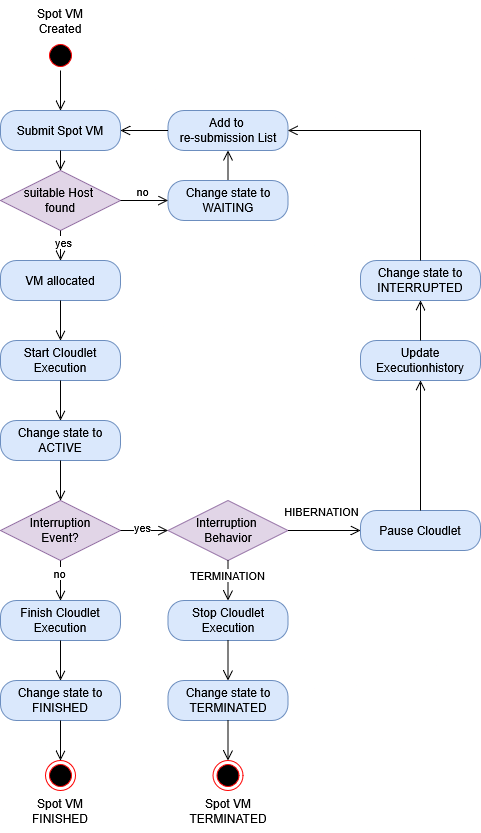}
    \caption{Spot instance lifecycle state transitions.}
    \label{fig:spot_lifecycle_diagram}
\end{figure}

            To capture runtime behavior, the extension enriches the default VM lifecycle with configurable interruption events and recovery mechanisms. 
            Figure~\ref{fig:spot_lifecycle_diagram} depicts the lifecycle of a spot instance, including transitions for termination and hibernation. CloudSim Plus, by default, discards unfulfilled VM requests; the extension instead models persistent requests and introduces resubmission after hibernation.
            
            CloudSim Plus, by default, supports dynamic VM arrivals but immediately discards requests that cannot be fulfilled due to insufficient resources. The spot instance extension overrides this behavior by modeling persistent VM requests and enriching the lifecycle with configurable interruption events and recovery mechanisms. These enhancements allow for a better representation of the uncertainty and volatility associated with spot pricing models in commercial cloud markets.
            
            Given the study’s focus on spot instance behavior, several new features were developed to simulate their lifecycle more realistically. One critical addition is the support for dynamic termination and hibernation. When an on-demand instance request cannot be fulfilled due to capacity constraints, the system attempts to free up resources by interrupting spot instances. The selection of which host to target for interruption depends on the active VM allocation policy.
            
            Each spot instance can be configured with an interruption behavior: either \textit{termination} or \textit{hibernation}. In the case of termination, the instance is immediately destroyed. If hibernation is selected, the instance is temporarily removed from the host, and its associated cloudlets are paused. Once sufficient capacity becomes available again, the spot instance is reallocated to a host, and its cloudlets resume execution.
            
            The implementation also includes tracking mechanisms for spot instance execution history. Each period of activity is recorded, allowing for the calculation of average interruption times and other metrics. To support realistic and configurable simulation behavior, the following time-based parameters were introduced:
            
            \begin{itemize}
                \item \textbf{Waiting Time:} Defines how long a persistent request remains active before being discarded.
                \item \textbf{Hibernation Time:} Specifies the maximum duration a spot instance may remain in hibernation before being terminated.
                \item \textbf{Minimum Running Time:} Ensures that spot instances cannot be interrupted due to capacity constraints until they have been running for a specified minimum duration.
                \item \textbf{Warning Time:} Defines the delay between an interruption signal and the actual termination, simulating a grace period.
            \end{itemize}
            
\subsection{Spot Behavior Extension for CloudSim Plus}

To support the simulation of spot instances, several key features were added. The primary enhancement is the \emph{differentiation of virtual machine types}, specifically the addition of on-demand and spot instances. Spot instances include advanced behaviors such as automatic \emph{termination or hibernation} when resources are needed for on-demand instances. Hibernated spot instances can be automatically \emph{resumed} once sufficient resources become available. The interruption behavior (termination or hibernation) and resubmission time window can be configured individually for each spot instance.
    
Additionally, \emph{persistent allocation requests} were introduced. In the standard CloudSim Plus framework, if a virtual machine cannot be allocated immediately, the request is discarded. In contrast, the extended implementation allows users to define a \emph{waiting time} during which the request remains active. If sufficient resources become available within this window, the virtual machine will be instantiated.

\subsection{Newly Introduced Classes}

To implement dynamic spot instance management, several new classes were introduced as extensions to CloudSim Plus. Each class integrates with existing framework abstractions but provides additional functionality tailored to spot market behavior.

\paragraph{DataCenterBrokerDynamic.} 
Extends \class{DatacenterBrokerAbstract} and introduces mechanisms for resubmitting VMs that fail to be created or are interrupted. This ensures that workloads associated with interrupted spot instances can be transparently retried, maintaining higher system utilization. A central feature is the \func{resubmittingList}, which tracks VMs awaiting reallocation.

\paragraph{DynamicAllocation.} 
Implements the core logic for allocating VMs while accounting for the preemptible nature of spot instances. Extending \class{VmAllocationPolicyAbstract}, it can free host capacity by deallocating or hibernating spot VMs, depending on their interruption settings. Methods such as \class{spotAllocation} and \func{terminationBehavior} model realistic cloud provider behavior when resolving resource contention.

\paragraph{DynamicVm.} 
Provides an abstract base for extended VM functionality. In addition to standard resource parameters, it supports persistent requests, configurable waiting times, and explicit VM states (e.g., \code{WAITING}, \code{INTERRUPTED}, \code{TERMINATED}). These additions allow simulations to capture the lifecycle of VMs under volatile spot market conditions.

\paragraph{OnDemandInstance and SpotInstance.} 
Two concrete subclasses of \class{DynamicVm}. On-demand instances behave as standard non-interruptible VMs. Spot instances, in contrast, include parameters such as interruption behavior (hibernate or terminate), minimum runtime guarantees, and hibernation timeouts, enabling more accurate modeling of provider-specific policies.

\paragraph{ExecutionHistory.} 
A helper class that records execution intervals of spot instances, including host, start, and stop times. These records enable the calculation of average interruption times and support detailed post-simulation analysis.

\paragraph{Table Builders.} 
Custom classes (\class{DynamicVmTableBuilder}, \class{SpotVmTableBuilder}, \class{ExecutionTableBuilder}) were added to extend CloudSim Plus’s reporting capabilities. They summarize VM lifecycle data, interruption statistics, and execution histories, and support export to CSV/JSON for external analysis.

\subsection{Modified Classes}

To maintain modularity, changes to existing CloudSim Plus classes were kept minimal. The \class{TableBuilderAbstract} class was extended with CSV export functionality, and small adjustments were made for the \emph{Google TraceReader} example (Section~\ref{trace}). No modifications were required in the simulation kernel, ensuring compatibility with the upstream framework.

\subsection{Reflections on Design Choices}

The design of the spot instance extension was shaped by both practical constraints and methodological considerations. Developing a simulation framework from scratch was not feasible within the scope of this work, as it would have required significant time and effort to replicate core infrastructure modeling. Instead, the decision was made to extend an existing simulator. CloudSim Plus was chosen because it is actively maintained, resolves several limitations of its predecessor CloudSim, and provides features such as dynamic workload support, runtime entity creation, and flexible event listeners \cite{silva2017cloudsim}. These characteristics made it a suitable foundation for extending cloud simulation capabilities with spot instance support.  

Overall, the design choices reflect a pragmatic balance between fidelity and feasibility. By extending CloudSim Plus rather than creating a new simulator, this work ensured compatibility with existing research tools while enabling the introduction of new functionality for modeling spot instance behavior. The use of Java further supported extensibility and long-term maintainability, aligning with the broader research goal of providing a reusable and adaptable simulation platform.  

\section{Allocation of Virtual Machines in Dynamic Spot Markets} \label{algorithm}

        The efficient placement of virtual machines is a central challenge in cloud computing, particularly when simulating dynamic environments such as those involving spot instances. VM allocation algorithms play a critical role in determining which physical host is assigned to a given VM request, directly impacting resource utilization, energy efficiency, and service continuity. 
        
        This chapter addresses the second research question posed in Chapter~\ref{question}, which investigates how VM placement strategies can be adapted to reduce spot instance interruptions and improve overall resource efficiency in dynamic cloud environments.
        
        The implemented approach builds on and extends the HLEM-VMP algorithm, which incorporates resource load balancing and energy-aware decisions into the placement process. The chapter is structured as follows: Section~\ref{algorithm-hlemvmp} introduces the conceptual foundation of the HLEM-VMP algorithm. Section~\ref{hlemvmp} discusses the implementation of the algorithm within the CloudSim Plus framework. Finally, Section~\ref{algorithm-adjustment} describes enhancements introduced to reduce spot instance interruptions and improve fairness across physical hosts.
    
        \subsection{HLEM-VMP: A Heuristic Allocation Algorithm} \label{algorithm-hlemvmp}
    
            To support the placement of virtual machines in dynamic cloud environments a VM allocation algorithm was implemented. Care was taken to extend the CloudSim Plus framework without modifying its core classes. This design ensures compatibility with future versions and maintains the modularity and extendability of the original framework. For instance, it remains possible to introduce additional allocation algorithms for both cloudlets and virtual machines.
            
            In selecting a suitable allocation algorithm, the dynamic nature of cloud workloads, particularly when simulating spot instances, was a key consideration. Traditional metaheuristic algorithms, such as Genetic Algorithms or Swarm Intelligence, are often employed to solve static bin packing or resource allocation problems. However, in this study, the workloads are time-dependent: virtual machines are dynamically created and terminated, making this a \textit{temporal bin packing problem}.
            
            Applying traditional GA-based approaches in this context would significantly increase complexity, as the algorithm must not only consider multi-dimensional resource constraints (CPU, RAM, etc.) but also the exact timing and duration of VM placement. The evolving simulation environment—with spot instance interruptions, resubmissions, and changing host availability—further complicates the use of such heuristics.
            
            Instead, a heuristic load- and energy-aware VM placement algorithm was implemented to better address the requirements of a dynamic scheduling environment. HLEM-VMP aims to balance the load across physical hosts and reduce unnecessary energy consumption, while accounting for constraints introduced by spot instance behavior. This enables faster decision-making and greater flexibility during simulation execution.
            
            The HLEM-VMP algorithm consists of three phases: host filtering, host load evaluation, and host selection \cite{jing2024hlemvmp}. In the host filtering phase, only hosts with sufficient free resources to meet the VM's requirements are selected. All resource types—CPU, memory, bandwidth, and storage—are considered. Additionally, unlike the original algorithm, this implementation checks the potential capacity of hosts if active spot instances were to be deallocated. This extension is necessary to accommodate the specific requirements of spot instance management.
            
            To avoid placing VMs with similar workloads on the same host, the algorithm introduces the $\text{RsDiff}$ value (Equation~\ref{eq:rsdiff}). This is calculated by subtracting the current CPU utilization of the physical host $U^1_i(t)$ from the CPU request of the VM $R^1_j(t)$. Only hosts for which $\text{RsDiff}$ exceeds a predefined threshold $\text{Thr}_{\text{cpu}}$ are considered for VM placement.
            
            \begin{equation}
            \text{RsDiff} = R^1_j(t) - U^1_i(t) * Rc
            \label{eq:rsdiff}
            \end{equation}
            
            \begin{equation}
            \text{RsDiff} > \text{Thr}_{\text{cpu}}
            \label{eq:rsdiff-threshold}
            \end{equation}
            
            The second phase, host load evaluation, involves computing weights for each resource to assess host suitability. First, the standardized available capacity $C^d_i(t)$ of each resource type $d$ on each host $i$ is normalized. Next, the proportion of each resource available on host $i$ relative to all hosts is computed.
            
            \begin{equation}
            _i(t) = \frac{C^d_i(t) - \min \{ C^d_1(t), \ldots, C^d_n(t) \}}{\max \{ C^d_1(t), \ldots, C^d_n(t) \} - \min \{ C^d_1(t), \ldots, C^d_n(t) \}}
            \label{eq:normalized-capacity}
            \end{equation}
            
            \begin{equation}
            p_{id} = \frac{C_i^d(t)}{\sum_{i=1}^{n} C_i^d(t)}, \quad (i = 1, \ldots, n;\ d = 1, \ldots, D)
            \label{eq:proportion}
            \end{equation}
            
            The entropy $e^d$ for each resource is then calculated based on $p_{id}$. Entropy quantifies the uncertainty or distribution imbalance of the resource:
            
            \begin{equation}
            e^d = -k \sum_{i=1}^{n} p_{id} \ln(p_{id})
            \end{equation}
            
            The constant $k$ is defined as:
            
            \begin{equation}
            k = \frac{1}{\ln(n)}
            \end{equation}
            
            Using entropy, the variation factor $g^d$ for each resource is computed. This measures the degree to which the resource influences the overall host evaluation. Weights for each resource are then derived by normalizing the variation factors.
            
            \begin{equation}
            g^d = 1 - e^d
            \end{equation}
            
            \begin{equation}
            w^d = \frac{g^d}{\sum_{d=1}^{D} g^d}
            \label{eq:weights}
            \end{equation}
            
            Finally, the overall host score $\text{HS}_i(t)$ is computed by summing the product of weights and the normalized capacities for each resource:
            
            \begin{equation}
            \text{HS}_i(t) = \sum_{d=1}^{D} w^d \cdot C_i^d(t)
            \label{eq:hostscore}
            \end{equation}
            
            The third and final phase is host selection. Hosts are sorted based on their computed scores, and starting from the host with the best score, each is evaluated for power consumption. If adding the VM does not exceed a predefined energy threshold, the VM is placed on that host \cite{jing2024hlemvmp}. In this implementation, energy consumption checks were omitted for simplicity, and the VM is allocated to the host with the highest score.
            
            The corresponding pseudocode implementation of the HLEM-VMP algorithm is provided in Algorithm~\ref{alg:helm-vpm}.

\begin{algorithm}[H]
\caption{HLEM-VMP VM Placement Strategy}
\label{alg:helm-vpm}
\footnotesize
\begin{algorithmic}[1]
\Require Hosts \code{PHs}, Virtual Machines \code{VMs}
\Ensure Placement of VMs on hosts
\State \textbf{Init:} \code{PHCandidateList} $\gets \emptyset$, $Thr_{cpu} \gets 0$, $R_c \gets 0.95$
\For{each $V_j$ in \code{VMs}}
    \State \code{PHCandidateList} $\gets$ \code{FilterPH(PHs, $V_j$)}
    \State \code{PHCandidateListClrSpot} $\gets$ \code{FilterPHWithSpotClr(PHs, $V_j$)}
    \If{$|\code{PHCandidateList}| > 0$}
        \State \code{HSlist} $\gets$ \code{CalPHScore(PHCandidateList)}
        \State Sort \code{PHCandidateList} by \code{HSlist(i)}
    \Else
        \State \code{HSlist} $\gets$ \code{CalPHScore(PHCandidateListClrSpot)}
        \State Sort \code{PHCandidateListClrSpot} by \code{HSlist(i)}
    \EndIf
\EndFor
\end{algorithmic}
\end{algorithm}
            
            Implementation specifics and performance evaluations of the HLEM-VMP algorithm are detailed in Sections~\ref{hlemvmp} and~\ref{algorithm-comparison}.\textbf{}
    
    \subsection{Implementation of the HLEM-VMP algorithm} \label{hlemvmp}
    
        This section introduces the HLEM-VMP (Heuristic Load and Energy-aware VM Placement) algorithm, which was implemented as a custom VM allocation policy in CloudSim Plus. The purpose of this algorithm is to optimize the placement of virtual machines with respect to load balancing and energy efficiency. The placement logic was implemented in the \class{DynamicAllocationHLEM} class, which extends the functionality of the \class{DynamicAllocation} class introduced in the previous subsection.
        
        \begin{itemize}
            \item \textbf{resourceCarryingFactor:} The \( R_c \) value used in the RsDiff calculation, as shown in Equation~\ref{eq:rsdiff}. The default value is 0.95.
            \item \textbf{threshold:} The threshold used for host filtering, as defined in Equation~\ref{eq:rsdiff-threshold}. The default value is 0.
        \end{itemize}
        
        \textbf{Methods:}
        
        \begin{itemize}
            \item \underline{defaultFindHostForVm:} This method contains the main loop of the algorithm. It iterates over all hosts, checks their suitability, and computes the RsDiff value for suitable hosts. For the filtered hosts, the \func{hostEvaluation} method is subsequently called.
            
            \item \underline{hostEvaluation:} First, the method \func{calculateCapacityVariables} is called to perform the initial two steps of the algorithm. Then, the entropy \( e^d \), the factor of variation \( g^d \), the weights for each resource \( w^d \), and finally the host scores \( \text{HS}_i(t) \) based on the normalized capacities are computed.
            
            \item \underline{calculateCapacityVariables:} This method calculates the standardized available capacity of each host \( C^d_i(t) \), as defined in Equation~\ref{eq:normalized-capacity}, as well as the proportional capacity \( p_{id} \), defined in Equation~\ref{eq:proportion}.
        \end{itemize}
    
    \subsection{Enhancing HLEM-VMP for Spot Instance Reliability} \label{algorithm-adjustment}
        
        While HLEM-VMP provides a heuristic-based VM placement strategy, further refinements were introduced to better align with the simulation goals—specifically, minimizing spot instance interruptions and promoting fairer host utilization. This section describes the adjustments made and the rationale behind them.
        
        To reduce the frequency of spot instance interruptions, the placement strategy aims to distribute spot instances more evenly across physical hosts. This is achieved by calculating the \emph{Spot Load} (\( \text{SL}_i(t) \)), which reflects the proportional resource usage of spot instances on a given host. The Spot Load is computed as the weighted sum over all resource types, where the weights are determined as described in Equation~\ref{eq:weights}:
        
        \begin{equation}
        \text{SL}_i(t) = \sum_{d=1}^{D} w^{(d)} \cdot \frac{C^{\text{spot}, d}_i(t)}{C^{\text{total}, d}_i(t)}
        \label{eq:spotload}
        \end{equation}
        
        This Spot Load value is scaled by a tunable parameter \( \alpha \), which defines its influence on host selection. The result is used to adjust the original Host Score (\( \text{HS}_i(t) \)), yielding an \emph{Adjusted Host Score} (\( \text{AHS}_i(t) \)) as follows:
        
        \begin{equation}
        \text{AHS}_i(t) = \text{HS}_i(t) \cdot (1 + \alpha \cdot \text{SL}_i(t))
        \label{eq:adjustedscore}
        \end{equation}
        
        To implement this logic, the original \class{DynamicAllocationHLEM} class was duplicated and modified, resulting in the new \class{DynamicAllocationHLEMAdjusted} class. Key methods were updated as described below: \\
        
        \textbf{Adjusted Methods:}
        \begin{itemize}
            \item \underline{calculateCapacityVariables:} After computing the proportional resource usage, the spot-specific capacity usage is calculated for each resource dimension (e.g., CPU, RAM, Bandwidth, Storage).
            
            \item \underline{hostEvaluation:} Before computing the original host scores \( \text{HS}_i(t) \), the Spot Load \( \text{SL}_i(t) \) is determined using the resource usage values and weights \( w^{(d)} \) from Equation~\ref{eq:spotload}. Once the base host scores are computed, they are adjusted according to Equation~\ref{eq:adjustedscore} by applying the Spot Load penalty factor \( \alpha \).
        \end{itemize}

    \section{Evaluation and Analysis} \label{evaluation}
    
        This section evaluates the developed extension for CloudSim Plus by defining relevant evaluation criteria and testing the implementation in different scenarios. The evaluation focuses on demonstrating the functionality of the extended features, especially the behavior of Spot Instances. To this end, two custom test cases were created. Additionally, real-world data from the Google Cluster Trace is used to assess the implementation under realistic workloads. The goal is to investigate how the simulation environment responds to dynamic instance behavior and cloud market conditions.

\subsection{Simulation Setup and Execution}

This section describes the steps required to implement and execute a minimal example of the developed extension. In this example, a single datacenter with one host is created to demonstrate the lifecycle of both on-demand and spot instances.

To begin, clone the Git repository containing the extension \footnote{The complete source code for the implementation is available at \url{https://github.com/CGoldi/cloudsimplus6-spot-instance.git}.}. A set of ready-to-use test cases is available and discussed in Section~\ref{testcases}.

        \paragraph{Initialize Simulation Components.}
        
        The following variables are required to initialize the simulation:

\begin{lstlisting}[language=Java, caption={Simulation and Broker Declarations}, label={lst:simulation-vars}]
private final CloudSim simulation;
private final DatacenterBrokerDynamic broker0;
private final List<DatacenterBroker> brokerList = new ArrayList<>();
\end{lstlisting}

        First, create the simulation instance using the \class{CloudSim} class. A minimum time between events and a simulation termination time are set.

\begin{lstlisting}[language=Java, caption={Simulation Setup}, label={lst:simulation-init}]
simulation = new CloudSim(0.5);
simulation.terminateAt(70);
\end{lstlisting}

        \paragraph{Create Host and Datacenter.}

        A list containing a single host is created. The host is initialized using the \class{HostDynamic} class with a specified amount of RAM, bandwidth, storage, and processing elements (PEs):

\begin{lstlisting}[language=Java, caption={Create Host}, label={lst:create-host}]
final List<Pe> peList = List.of(new PeSimple(1000), new PeSimple(1000));
Host host = new HostDynamic(2048, 10000, 1000000, peList);
\end{lstlisting}

        The datacenter is instantiated using the \class{DatacenterSimple} class. The host list and a custom allocation policy, \class{DynamicAllocationHLEM} (based on HELM-VPM), are passed as parameters.

\begin{lstlisting}[language=Java, caption={Create Datacenter}, label={lst:create-dc}]
final DynamicAllocationHLEM allocationPolicy = new DynamicAllocationHLEM();
Datacenter datacenter0 = new DatacenterSimple(simulation, List.of(host), allocationPolicy);
datacenter0.setSchedulingInterval(1);
\end{lstlisting}

        \paragraph{Create Broker.}
        
        The simulation broker, based on the \class{DatacenterBrokerDynamic} class, is used to submit VMs and cloudlets to the datacenter.

\begin{lstlisting}[language=Java, caption={Create Broker}, label={lst:create-broker}]
broker0 = new DatacenterBrokerDynamic(simulation);
brokerList.add(broker0);
broker0.setShutdownWhenIdle(false);
broker0.setVmDestructionDelay(1);
\end{lstlisting}

        \paragraph{Define Virtual Machines.}
        
        Spot instances are created using the \class{SpotInstance} class. Configuration parameters such as RAM, bandwidth, storage size, and interruption behavior (e.g., \code{HIBERNATE}) are set.

\begin{lstlisting}[language=Java, caption={Create Spot VM}, label={lst:create-spotvm}]
final SpotInstance spotvm = new SpotInstance(1000, 2, true);
spotvm.setRam(512);
spotvm.setBw(1000);
spotvm.setSize(10000);
spotvm.setInterruptionBehavior(SpotInstance.InterruptionBehavior.HIBERNATE);
spotvm.addOnHostDeallocationListener(this::onHostDeallocationListener);
\end{lstlisting}

        On-demand instances use the \class{OnDemandInstance} class and may be delayed using \func{setSubmissionDelay}:

\begin{lstlisting}[language=Java, caption={Create On-Demand VM}, label={lst:create-ondemandvm}]
final OnDemandInstance ondemandvm = new OnDemandInstance(1000, 2, true);
ondemandvm.setRam(512);
ondemandvm.setBw(1000);
ondemandvm.setSize(10000);
ondemandvm.setSubmissionDelay(10);
ondemandvm.addOnHostDeallocationListener(this::onHostDeallocationListener);
\end{lstlisting}

        \paragraph{Create Cloudlets.}
        
        Each VM must be associated with a cloudlet that determines its execution workload during the simulation:

\begin{lstlisting}[language=Java, caption={Create Cloudlet}, label={lst:create-cloudlet}]
Cloudlet cloudletSpot = new CloudletSimple(1, 20000, 1);
cloudletSpot.setFileSize(300);
cloudletSpot.setOutputSize(300);
cloudletSpot.setUtilizationModel(new UtilizationModelFull());
cloudletSpot.setVm(spotvm);
\end{lstlisting}

        \paragraph{Submit VMs and Cloudlets.}
        
        VMs and cloudlets are submitted to the broker for scheduling and execution:

\begin{lstlisting}[language=Java, caption={Submit VMs and Cloudlets}, label={lst:submit-vms}]
broker0.submitVm(spotvm);
broker0.submitCloudlet(cloudletSpot);
broker0.submitVm(ondemandvm);
broker0.submitCloudlet(cloudletOnDemand);
\end{lstlisting}

        \paragraph{Simulation Loop and Execution.}
        
        The simulation must update the processing of all active VMs on each clock tick. This is done by registering an event listener:

\begin{lstlisting}[language=Java, caption={Update VM Processing}, label={lst:update-vm}]
private void updateProcessingforVms(EventInfo eventInfo) {
  for (DatacenterBroker broker : brokerList) {
    for (Vm vm : broker.getVmExecList()) {
      vm.updateProcessing(simulation.clock(), vm.getHost().getVmScheduler().getAllocatedMips(vm));
    }
  }
}
\end{lstlisting}

        Simulation execution begins with the following command:

\begin{lstlisting}[language=Java, caption={Start Simulation}, label={lst:start-simulation}]
simulation.addOnClockTickListener(this::updateProcessingforVms);
simulation.start();
\end{lstlisting}

        \paragraph{Output Results.}
        
        Output tables are generated using the provided builders. These summarize VM execution and scheduling behavior:

\begin{lstlisting}[language=Java, caption={Build Output Tables}, label={lst:output}]
new DynamicVmTableBuilder(finishedVms).build();

List<SpotInstance> finishedSpot = new ArrayList<>();
for (DynamicVm vm : finishedVms) {
  if (vm instanceof SpotInstance) {
    ((SpotInstance) vm).calculateAverageInterruptionTime();
    finishedSpot.add((SpotInstance) vm);
  }
}

new SpotVmTableBuilder(finishedSpot).build();
\end{lstlisting}
        
        \subsection{Implementation Test Cases} \label{testcases}
        
            To illustrate how the new features integrate with CloudSim Plus and how they behave during simulation, two test cases were developed: \\ \class{RandomlyGeneratedInstances} and \class{RestartingInterruptedSpot}.
            
            \paragraph{RandomlyGeneratedInstances.}
            
            This scenario demonstrates the automatic interruption and termination of Spot Instances. The simulation setup includes the creation of a simulation object, a datacenter with hosts, a datacenter broker, and a set of initial virtual machines and cloudlets. These VMs and cloudlets are submitted to the broker.
            
\begin{figure}[!t]
                \centering
                \includegraphics[width=\linewidth]{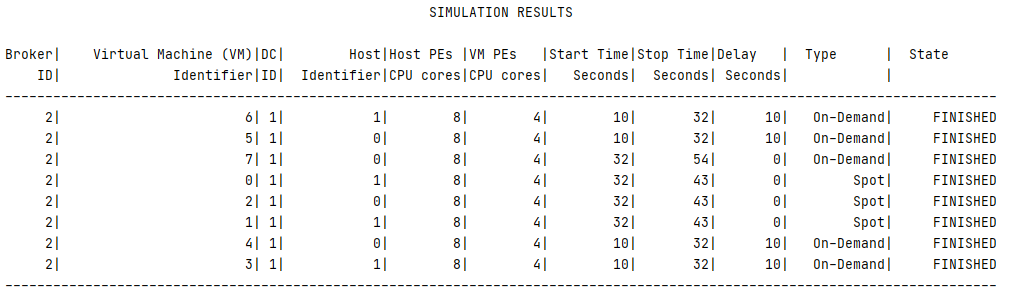}
                \caption{Example Output -- \class{RestartingInterruptedSpot} (DynamicVmTableBuilder)}
                \label{fig:output2_1}
            \end{figure}

            Additionally, a \func{clockTickListener} is added, which dynamically generates new VM instances and associated cloudlets during simulation runtime. These VMs may be either On-Demand or Spot Instances. If insufficient capacity is available to allocate an On-Demand instance, running Spot Instances will be terminated to free resources. The output of this simulation, shown in Figure~\ref{fig:output2_1}, presents a list of VMs executed during the simulation, including their final states. Interrupted Spot Instances appear with the status \code{TERMINATED}. Portions of the code were adapted from the examples in the CloudSim Plus repository.
            
            \paragraph{RestartingInterruptedSpot.}
            
            This test case demonstrates persistent request behavior and the resubmission of interrupted Spot Instances. The simulation setup is similar to the previous example, with the addition that Spot Instances are created first and On-Demand Instances are added with a delay of 10 seconds.
            
            To enable automatic resubmission, Spot Instances are configured with persistent requests, a maximum waiting time, and a hibernation time limit. \\ An \func{onHostDeallocationListener} is used to trigger re-submission whenever a VM completes execution on any host. Alternatively, a \func{clockTickListener} could be used for periodic checks.
            
\begin{figure}[!t]
                \centering
                \includegraphics[width=\linewidth]{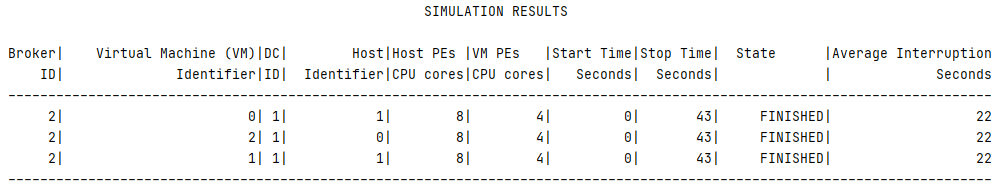}
                \caption{Example Output -- \class{RestartingInterruptedSpot} (SpotVmTableBuilder)}
                \label{fig:output2_2}
            \end{figure}

            During the simulation, Spot Instances are preempted by On-Demand Instances and later resumed when resources become available. The output is shown in Figures~\ref{fig:output2_1} and~\ref{fig:output2_2}. The first figure displays all VMs, while the second focuses on Spot Instances and includes their average interruption times.
            
\subsection{Google Cluster Trace} \label{trace}

To evaluate the implementation under realistic conditions, the spot instance extensions were tested with the Google Cluster Trace. CloudSim Plus provides a reader for this dataset, which was extended to integrate spot functionality.

This section briefly introduces Google’s Borg cluster manager and the 2011 trace release, followed by a description of the simulation process and results.

\paragraph{Google's Cluster Manager – Borg.}
Borg is Google’s internal cluster management system, orchestrating hundreds of thousands of tasks across datacenters. A cluster is composed of one or more \emph{cells}, each with around 10,000 machines of similar hardware configuration \cite{verma2015large}. Borg schedules both high-priority production services (e.g., Gmail, Docs), which dominate resource usage, and low-priority batch jobs, most of which are preemptible. At the time of the trace, workloads were executed directly on physical hosts, without virtualization.

\subsubsection{Google Cluster Data} \label{cluster11}
In 2011, Google released a 30-day trace from a production cluster with $\sim$12,600 machines \cite{reiss2011google,tirmazi2020borg}. The dataset records both machine events (e.g., addition, removal, resource updates) and job/task events (submission, scheduling, termination). In total, \emph{48.4M tasks} were submitted, with between \emph{97k and 223k tasks} active concurrently. The busiest host ran \emph{8,071 tasks} in total, with up to \emph{410} running in parallel.  

\begin{figure}[!t]
    \centering
    \includegraphics[width=\linewidth]{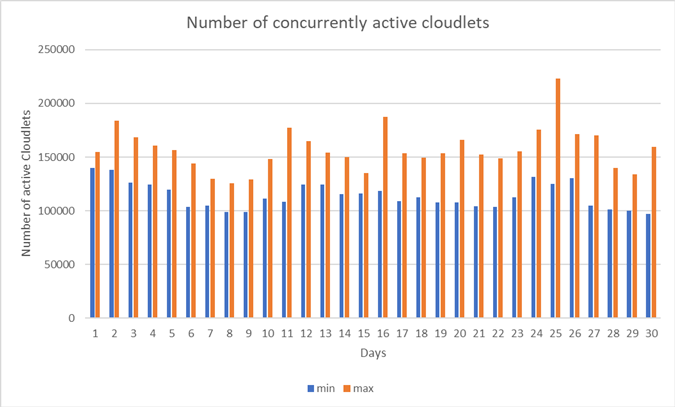}
    \caption{Maximum and minimum number of concurrently active tasks per day}
    \label{fig:active_tasks}
\end{figure}

\paragraph{Analysis.}
Figure~\ref{fig:active_tasks} shows daily ranges of concurrently active tasks, averaging between \emph{114k and 157k}. The workload was distributed across \emph{12,585 machines}. About \emph{1.7\%} of tasks lacked valid machine mappings and were excluded.

\begin{figure}[!t]
    \centering
    \includegraphics[width=\linewidth]{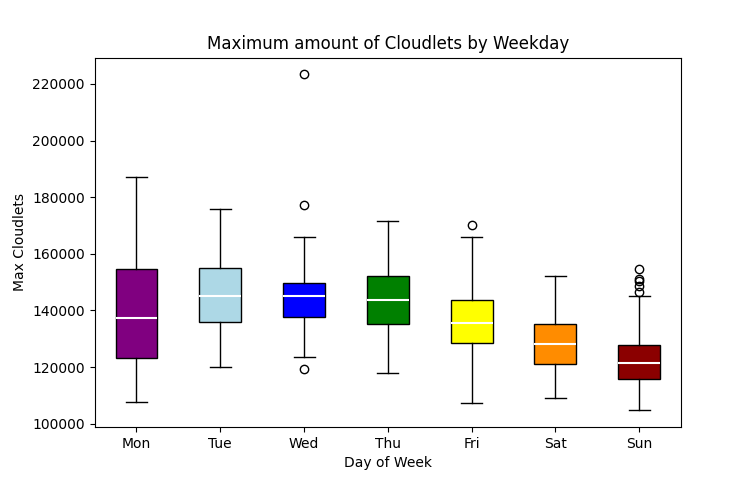}
    \caption{Daily distribution of maximum concurrently running cloudlets (hourly resolution)}
    \label{fig:boxplot_days}
\end{figure}

\begin{figure}[!t]
    \centering
    \includegraphics[width=\linewidth]{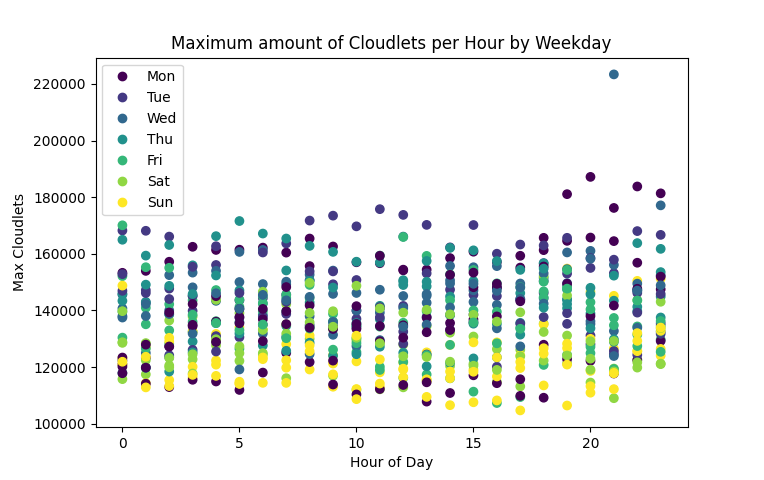}
    \caption{Maximum number of concurrently running cloudlets by hour of the day}
    \label{fig:scatter_hours}
\end{figure}

\paragraph{Simulation-Specific Data.}
Because Borg does not use VMs, task submissions were grouped into synthetic VMs by user and machine ID. This produced \emph{28.8M VMs}, with between \emph{76k and 129k} concurrently active. Figure~\ref{fig:boxplot_days} and Figure~\ref{fig:scatter_hours} illustrate daily and hourly variations in active cloudlets.

\subsubsection{Simulation with Cluster Data}
The simulation combines the \class{Machine Events} and \class{Task Events} tables. Machine events create and remove hosts, while task events generate VMs and cloudlets.

\paragraph{Trace Reader.}
The CloudSim Plus trace reader was extended to (i) bind tasks to machines during submission, (ii) use a \func{cloudletHashMap} for faster lookups, (iii) revise handling of \code{EVICT}/\code{FAIL} events, and (iv) defer delayed submissions to reduce overhead. Missing machine attributes were filled by replication, and missing task–machine mappings were resolved from later events.

\paragraph{Data Preparation.}
Incomplete CPU and RAM entries in the machine table were filled with values from other machines. Task events missing machine IDs were reconciled by checking subsequent events; unresolved entries were allocated by policy.

\subsection{Simulation Implementation \& Output}
The simulation integrates both host and task creation. In addition to trace workloads, \emph{200,000 spot instances} were injected, each with fixed durations (20/40 hours) to ensure completion despite interruptions. Outputs include CSV/JSON files of VMs, cloudlets, and spot instance interruption histories.

\subsubsection{Simulation Performance} \label{performance}
A one-day trace simulation required $\sim$1.5 days on a \emph{e2-highmem-4} VM (4 vCPUs, 32 GB RAM), with CPU and memory usage below 30\%. A two-day trace took $\sim$7 real days. Performance was constrained by cloudlet execution updates rather than allocation, suggesting parallelization as a future optimization.

\begin{figure}[!t]
    \centering \includegraphics[width=\linewidth]{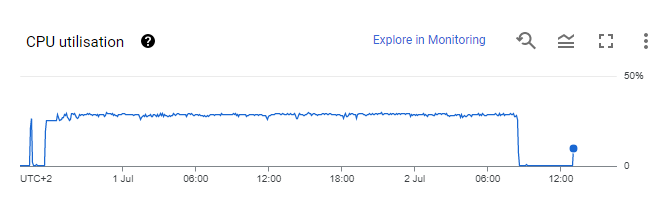}
    \caption{CPU utilization during one-day simulation}
    \label{fig:cpu_simulation}
\end{figure}

\begin{figure}[!t]
    \centering \includegraphics[width=\linewidth]{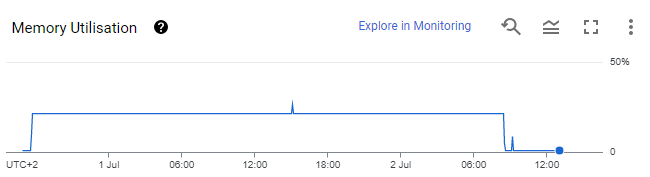}
    \caption{Memory utilization during one-day simulation}
    \label{fig:memory_simulation}
\end{figure}

\subsubsection{Final Simulation \& Results}
The final two-day simulation created \emph{2.38M trace VMs} and \emph{200k spot instances}.  

\textbf{Spot interruptions:} 16.5\% completed without interruption; 166,918 were interrupted; 92,554 redeployed (most once, a few up to three times). Average interruption time was $\sim$1,910 s (32 min), with a maximum of 7,711 s (2.1 h).  

\textbf{Completion:} 38.5\% of spot instances finished, including 43,878 after interruption; 123,040 were terminated.  

\begin{figure}[!t]
    \centering
    \includegraphics[width=\linewidth]{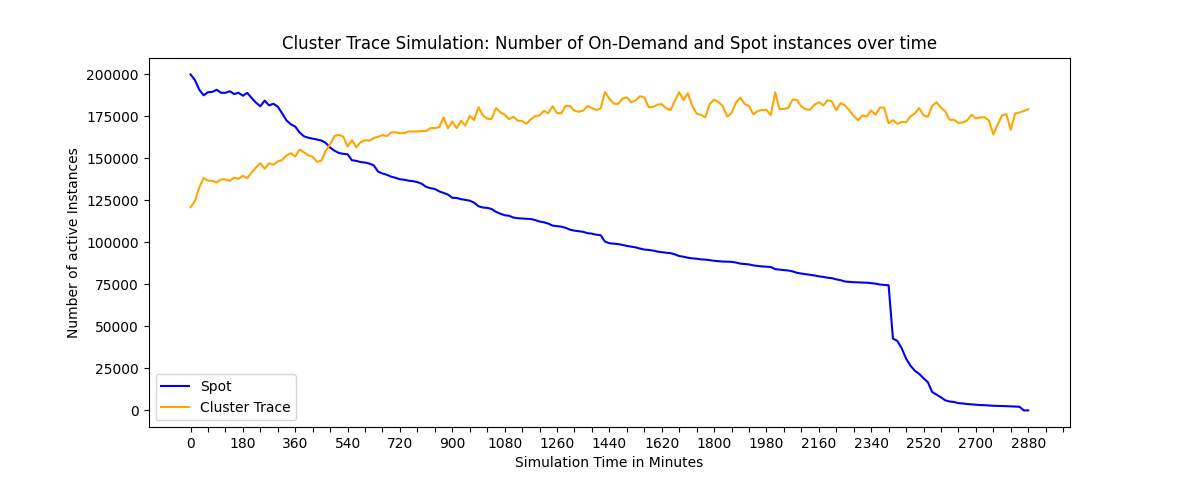}
    \caption{Cluster Trace Simulation: Active VM instances over time}
    \label{fig:active_instances}
\end{figure}

Figure~\ref{fig:active_instances} shows VM activity over time. Increases in trace load coincide with higher eviction rates, while dips allow temporary recovery of spot instances.

\subsection{Comparison of Allocation Algorithms} \label{algorithm-comparison}
            
            To evaluate the performance and effectiveness of the implemented HLEM-VMP algorithm and its adjusted version, a comparative analysis was conducted using a sample simulation. Additionally, both algorithms were compared with the First-Fit algorithm, which is already included in the CloudSim Plus framework.
            
            \subsubsection{Evaluation Criteria}
            
            The evaluation was based on the following key metrics:
            \begin{itemize}
                \item Number of spot instance interruptions
                \item Average interruption time of spot instances
            \end{itemize}
            
            \subsubsection{Simulation Setup}
            
            Table~\ref{tab:host_profiles} presents the host types used in the simulation. Four host configurations—\textit{small}, \textit{medium}, \textit{large}, and \textit{x-large}—were defined, with each successive type offering increased CPU, memory, bandwidth, and storage capacities. The simulation included 20 small, 30 medium, 30 large, and 20 x-large hosts.
            
            Virtual machines were defined using 10 distinct VM profiles, each with varying resource demands in terms of CPU, memory, bandwidth, and storage. These profiles are detailed in Table~\ref{tab:vm_profiles}. Randomized values were used for both VM submission delays and total execution times. The same randomized values were reused across all simulation runs to ensure consistency.
            
            A total of 2,000 VMs were submitted: 400 spot VMs and 600 on-demand VMs were submitted immediately, without delay. The remaining 1,000 VMs (50\% of the total) were submitted with randomized delays. All randomized variables, including instance type, VM profile, delay, and execution duration, were kept identical across all algorithms to enable a fair comparison.
            
            The scheduling algorithms evaluated in this section are the \emph{HLEM-VMP} (Section~\ref{hlemvmp}), the \emph{adjusted HLEM-VMP} (Section~\ref{algorithm-adjustment}), and the \emph{First-Fit} baseline provided by CloudSim Plus.
            
            \begin{table}[ht]
                \centering
                \caption{Host Types}
                \label{tab:host_profiles}
                \begin{tabular}{lcccc}
                    \toprule
                    \textbf{Size} & \textbf{CPU} & \textbf{Memory} & \textbf{Bandwidth} & \textbf{Storage} \\
                    \midrule
                    Small    & 8   & 16,384  & 5,000   & 200,000  \\
                    Medium   & 16  & 32,768  & 10,000  & 400,000  \\
                    Large    & 32  & 65,536  & 20,000  & 800,000  \\
                    X-Large  & 64  & 131,072 & 40,000  & 1,600,000\\
                    \bottomrule
                \end{tabular}
                \vspace{0.5em}
            \end{table}
            
            \begin{table}[ht]
                \centering
                \caption{VM Profiles}
                \label{tab:vm_profiles}
                \begin{tabular}{lccccc}
                    \toprule
                    \textbf{CPU} & \textbf{Memory} & \textbf{Bandwidth} & \textbf{Storage} & \textbf{Spot \#} & \textbf{On-Demand \#} \\
                    \midrule
                    1  & 1,024   & 100   & 10,000   & 31  & 160 \\
                    2  & 1,024   & 100   & 10,000   & 42  & 175 \\
                    1  & 2,048   & 200   & 20,000   & 36  & 168 \\
                    2  & 2,048   & 200   & 20,000   & 44  & 146 \\
                    4  & 2,048   & 200   & 20,000   & 40  & 158 \\
                    4  & 4,096   & 500   & 50,000   & 40  & 145 \\
                    6  & 4,096   & 500   & 50,000   & 36  & 170 \\
                    6  & 8,192   & 1,000 & 80,000   & 51  & 155 \\
                    8  & 8,192   & 1,000 & 80,000   & 33  & 162 \\
                    10 & 8,192   & 1,000 & 80,000   & 47  & 168 \\
                    \bottomrule
                \end{tabular}
                \vspace{0.5em}
            \end{table}

            \subsubsection{Results and Analysis}
            
\begin{figure}[!t]
                \centering
                \includegraphics[width=\linewidth]{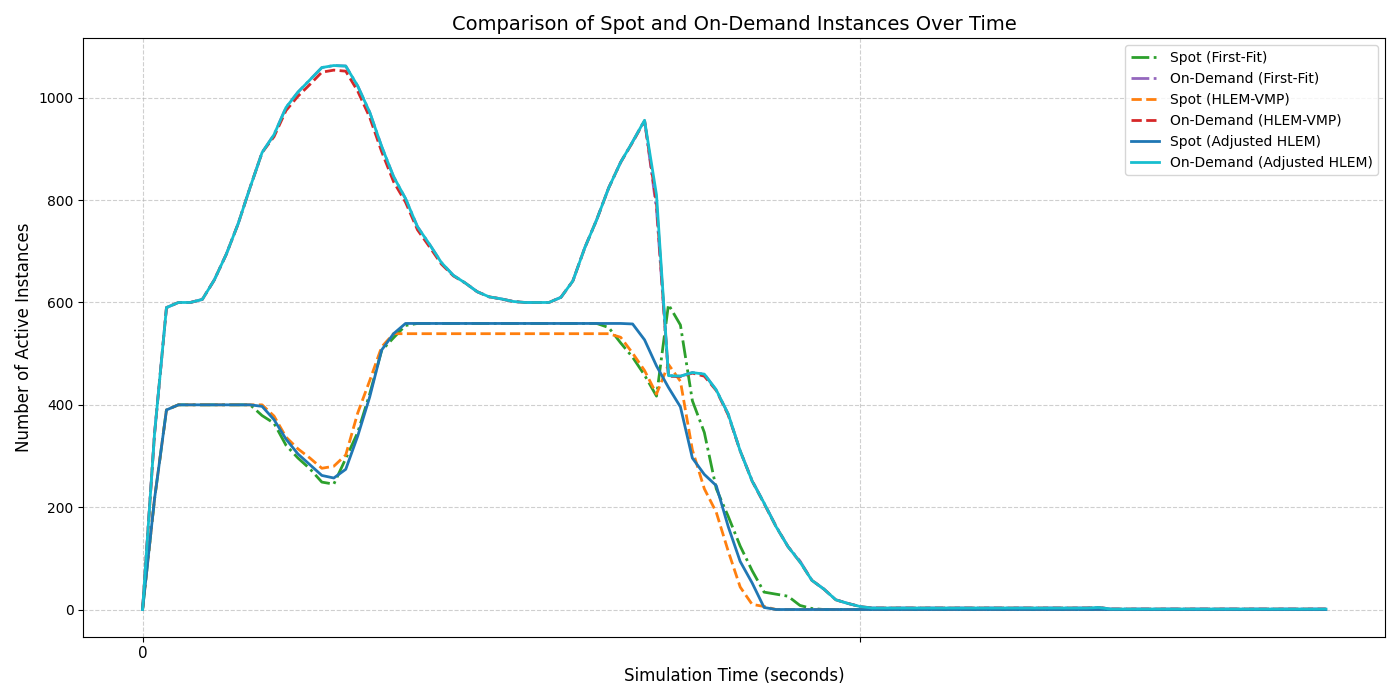}
                \caption{Active Instances Over Time}
                \label{fig:instances_over_time}
            \end{figure}

            This subsection presents the simulation results and highlights performance differences among the algorithms. Figure~\ref{fig:instances_over_time} displays the number of active spot and on-demand VMs over time. While on-demand activity remained nearly identical across all algorithms, differences emerged in the distribution of spot instances.

\begin{figure}[!t]
                \centering
                \includegraphics[width=\linewidth]{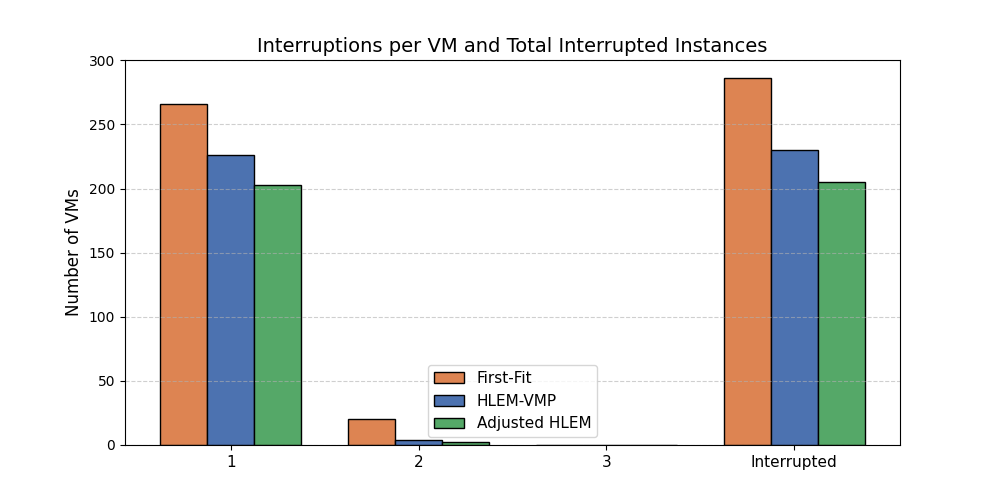}
                \caption{Total Spot Instance Interruptions}
                \label{fig:total_interruptions}
            \end{figure}

            Figure~\ref{fig:total_interruptions} shows the total number of spot instance interruptions. The First-Fit algorithm resulted in the most interruptions (286), followed by HLEM-VMP (230), and the adjusted HLEM-VMP performed best with only 205 interruptions.

\begin{figure}[!t]
                \centering
                \includegraphics[width=\linewidth]{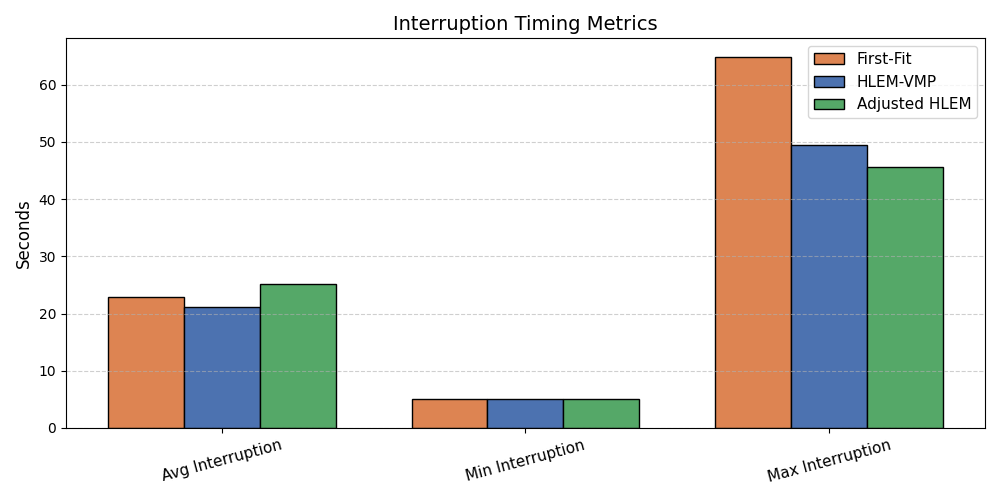}
                \caption{Spot Instance Interruption Durations}
                \label{fig:interruption_timings}
            \end{figure}

            In all scenarios, the maximum number of interruptions per VM was two. Average interruption times are shown in Figure~\ref{fig:interruption_timings}. HLEM-VMP achieved the shortest average interruption time (21.12 seconds), followed by First-Fit (22.81 seconds), while the adjusted HLEM-VMP had the highest (25.20 seconds). However, when comparing maximum interruption durations, the adjusted HLEM-VMP performed best (45.65 seconds), followed by HLEM-VMP (49.49 seconds), and First-Fit (64.87 seconds). Minimum interruption times were nearly identical across all algorithms.

            Interruption statistics were gathered using the \class{SpotVmTableBuilder} for aggregated interruption data and the \class{ExecutionTableBuilder} for execution timelines. These results confirm that allocation strategies can meaningfully affect spot instance behavior and demonstrate that adjustments to HLEM-VMP can reduce interruption frequency in dynamic cloud simulations.
            
    \subsection{Outlook: Interruption Frequency Data}
    
    The exact criteria cloud providers such as AWS use to decide which spot instances are interrupted remain undisclosed. This raises the question of whether certain instance attributes influence the likelihood of interruption. 
    
\begin{figure}[!t]
        \centering
        \includegraphics[width=\linewidth]{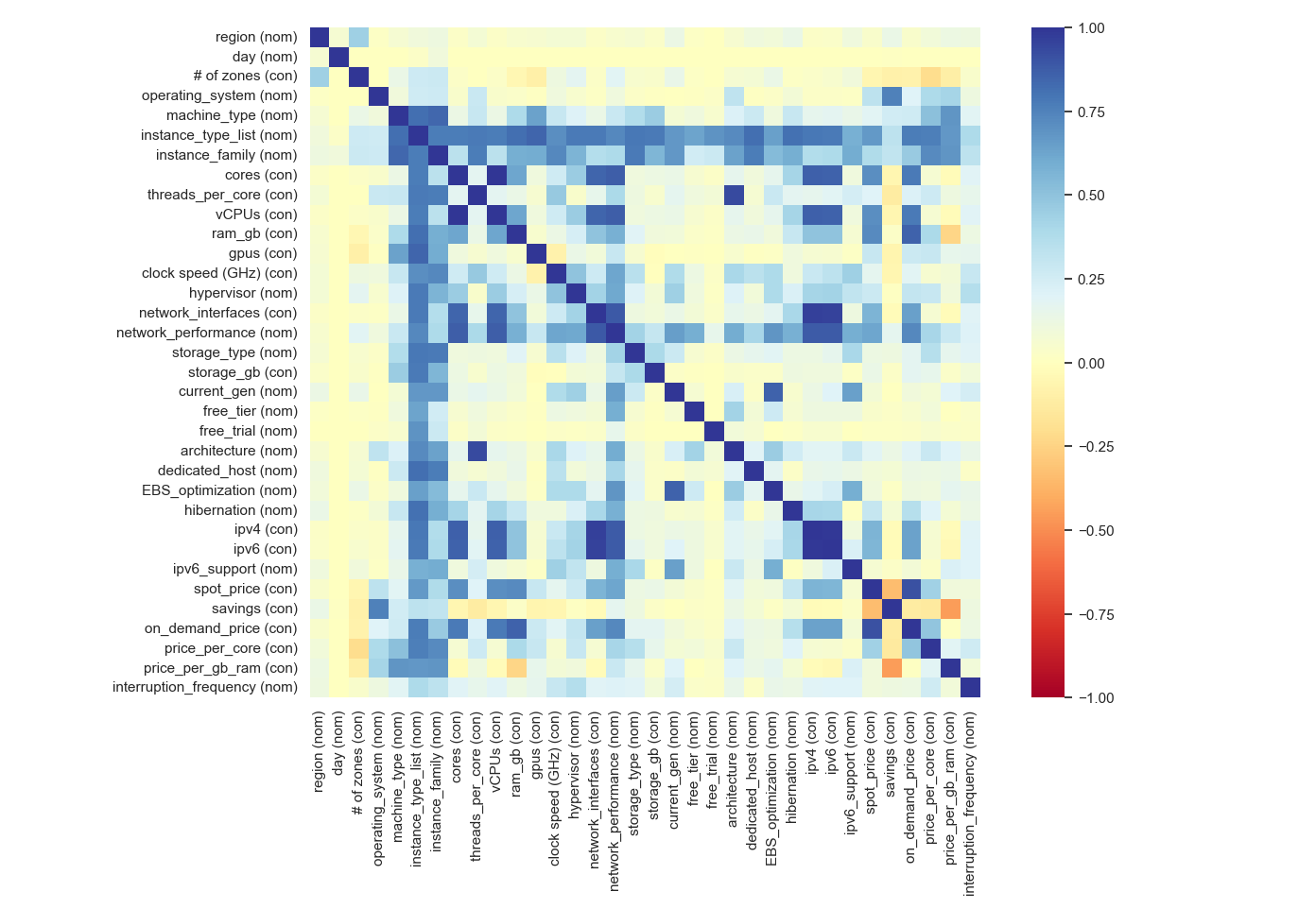}
        \caption{Correlation analysis of instance attributes and interruption frequency.}
        \label{fig:advisor}
    \end{figure}

    AWS provides the Spot Instance Advisor\footnote{\url{https://spot-bid-advisor.s3.amazonaws.com/spot-advisor-data.json}}, which categorizes interruption frequency into five ranges: $<$5\%, 5–10\%, 10–15\%, 15–20\%, and $>$20\%. While this dataset offers general insights, it lacks granularity, particularly in the highest category. Figure~\ref{fig:advisor} illustrates a correlation analysis based on this data.
    
    In addition to interruption frequency, the Spot Advisor dataset provides attributes such as instance category (general purpose, compute optimized), vCPU count, memory size, and expected cost savings. It is region-specific and distinguished by operating system (Linux/Unix, Windows). To extend the feature set, we integrated spot price data from the AWS Spot Price feed\footnote{\url{https://spot-price.s3.amazonaws.com/spot.js}} and metadata from AWS console exports, including availability zones, GPU count, generation, storage type, and on-demand pricing. In total, 389 instance types were collected, though not all are available in each region.
    
    The dataset was preprocessed for quantitative analysis: numeric fields were normalized, categorical features encoded, and derived attributes created (e.g., price per GB of RAM). Instances were also hierarchically grouped into categories, families, and exact types. This resulted in a comprehensive dataset for analyzing interruption frequency.
    
    \paragraph{Feature Analysis.}
    Correlation analysis was performed using the \texttt{dython.nominal} library, which applies suitable measures for mixed data: Theil’s U for nominal–nominal, correlation ratio ($\eta^2$) for numeric–categorical, and Pearson correlation for numeric–numeric features. The strongest associations with interruption frequency were:
    
    \begin{itemize}
        \item \textbf{Instance type:} 0.38
        \item \textbf{Instance family:} 0.33
        \item \textbf{Machine type:} 0.18
    \end{itemize}
    
    This suggests that more fine-grained classifications (e.g., exact type) carry stronger predictive power than broader categories. Features such as day, free\_tier, and dedicated\_host showed negligible correlation and were excluded.
    
    For future analysis, methods such as Mutual Information, tree-based feature importance (Random Forests), and ANOVA/Chi-square tests could be applied to capture non-linear dependencies and validate findings.
    
\section{Limitations}
\label{sec:limitations}

This work focuses on extending CloudSim Plus with spot instance support rather than developing a new simulation framework or conducting a full market or economic analysis.

Due to resource constraints (Google Cloud VM under free-tier), only a subset of cluster data was used for simulation. Select optimizations, such as omitting unnecessary runtime calculations, were applied to reduce computational load.

The interruption frequency analysis of AWS spot instances relied on static snapshot data from the Spot Instance Advisor. While sufficient for exploratory purposes, it does not capture temporal variations. Long-term time-series data collection was outside the scope of this work.

Advanced data science techniques were not applied; feature analysis was exploratory and aimed at identifying correlations relevant to simulation behavior rather than building predictive models.

The use of CloudSim Plus provided a mature and modular base but imposed structural constraints, as some core classes are not easily modifiable. These limitations were accepted to ensure framework compatibility and stability.

\section{Conclusion}
\label{sec:conclusion}

Over time, the scientific community has proposed various perspectives on the evolution of future cloud markets, highlighting economic considerations even within Grid computing~\cite{weishaupl2005business,schikuta2005business,schikuta2008workflow}, the computational paradigm that preceded Clouds. Trust and security are recognized as fundamental enablers of the digital economy, as they cultivate confidence among stakeholders, thereby encouraging individuals and organizations to participate in digital transactions and share sensitive information. In the absence of trust~\cite{weishaupl2006gset}, users may be reluctant to adopt digital platforms~\cite{fuerte2000meta}, while a lack of security~\cite{shaaban2019ontology} could expose the ecosystem to risks such as fraud, data breaches, and cyberattacks, ultimately compromising its integrity. Our prior and ongoing research in service classification~\cite{vinek2011classification,mach2012generic} and workflow composition~\cite{kofler2010user,stuermer2009building}, with a strong emphasis on economic aspects, has inspired our current investigation into integrating economic principles with secure and autonomous cloud management~\cite{pittl2018blockchain,pittl2016bazaar}, leading to the analysis of cloud marketspaces.

The widespread adoption of cloud platforms across diverse application domains has led to increasing interest in optimizing key aspects such as resource allocation, scheduling, energy efficiency, and virtual machine migration policies. However, conducting experiments on public cloud infrastructure is often associated with high operational costs and limited control over internal behaviors. Cloud simulation provides a cost-effective and scalable alternative, enabling researchers to evaluate new strategies under controlled conditions.

Among the available frameworks, CloudSim and CloudSim Plus are widely used in academic research. Despite their flexibility, these frameworks lack support for accurately modeling spot (or preemptible) instances—VMs offered at significant discounts by cloud providers to utilize spare capacity, but which can be interrupted at any time. Understanding the behavior of such instances is critical for developing effective, cost-saving strategies in Infrastructure-as-a-Service environments.

This work addressed this limitation by designing and implementing an extension to CloudSim Plus that enables realistic simulation of spot instance behavior. The extension models the full spot instance lifecycle, including the interruption of instances to free capacity for on-demand requests and configurable interruption strategies such as termination and hibernation. Furthermore, support for dynamic provisioning was introduced, allowing both spot and on-demand VMs to be resubmitted if their initial placement fails due to insufficient resources.

As a proof of concept, a simulation based on the Google Cluster Trace dataset was conducted. Rather than attempting to replicate the full trace, the simulation focused on demonstrating the feasibility and effectiveness of modeling dynamic spot instance behavior—specifically the hibernation and resumption of instances in response to resource availability. The scenario validates the ability of the extended framework to capture essential dynamics observed in real-world cloud environments.

To complement the modeling of spot instance behavior in CloudSim Plus, an allocation algorithm from existing literature—HLEM-VMP—was implemented. This heuristic-based algorithm utilizes host filtering and selection mechanisms to determine the most suitable host for a VM, with the objective of reducing SLA violations. It was chosen for its simplicity and its ability to support dynamic resource allocation without the overhead and complexity associated with meta-heuristic approaches. The primary input parameters for the algorithm are the resource demands of the virtual machines and the available capacities on the hosts.

The purpose of integrating this algorithm was to demonstrate how the developed extension enables the evaluation and comparison of different VM allocation strategies. To that end, the algorithm was further refined to reduce the number of interruptions by factoring in the proportion of host resources currently consumed by spot instances in the host selection score calculation.

The adjusted HLEM algorithm was evaluated against the original HLEM-VMP and the First-Fit strategy. Results demonstrated that the adjusted version could reduce the number and maximum duration of spot instance interruptions under certain conditions. However, this improvement came at the cost of a slightly increased average interruption time, highlighting a trade-off between frequency and duration of interruptions.

Currently, the selection of which spot instances to terminate during resource contention is determined in a non-deterministic manner, based solely on the VM list associated with a host. This introduces potential inefficiencies. Future research could explore the implementation of targeted deallocation strategies or predictive models to more intelligently select interruption candidates based on historical patterns, VM importance, or runtime behavior.

Additionally, a preliminary analysis was conducted to examine which instance characteristics correlate with spot instance interruption frequencies, using publicly available data from the AWS Spot Instance Advisor. Further research in this direction—supported by empirical data—could contribute to the refinement of simulation models and allocation strategies, ultimately improving the accuracy of spot instance behavior modeling, particularly in relation to termination patterns.

\emph{In summary}, this work contributes a meaningful extension to CloudSim Plus for simulating dynamic spot instance behavior and demonstrates its applicability in evaluating VM placement strategies. The proposed enhancements lay the groundwork for further research into cost-efficient, interruption-aware scheduling algorithms in cloud environments.


%

\ifCLASSOPTIONcaptionsoff
  \newpage
\fi

\IEEEtriggeratref{35}

\bibliographystyle{IEEEtran}
\bibliography{bibliography}

\end{document}